\documentclass[]{raa}
\usepackage{graphicx}
\usepackage{times}
\usepackage{natbib}
\usepackage{mathrsfs}
\usepackage{amsmath}
\usepackage{amssymb}
\usepackage{booktabs}
\usepackage{hyperref}

\usepackage{multirow}
\usepackage{threeparttable}

\begin{document}

   \title{FRB~171019: An event of binary neutron star merger?}
   \volnopage{Vol. 000 No.0, 000--000}      
   \setcounter{page}{1}          
   \author{Jinchen Jiang
      \inst{1,2}
   \and Weiyang Wang
      \inst{3,4}
   \and Rui Luo
      \inst{5}
   \and Shuang Du
      \inst{1}
   \and Xuelei Chen
      \inst{3}
   \and Kejia Lee
       \inst{1,6}
   \and Renxin Xu
       \inst{1,6}
   }

    \institute{%
           State Key Laboratory of Nuclear Physics and Technology, School of Physics, Peking University, Beijing 100871, China; {\em jiangjinchen@pku.edu.cn}\\
        \and
          National Astronomical Observatories, Chinese Academy of Sciences, Beijing, Beijing
          100101, China\\
       \and
         Key Laboratory of Computational Astrophysics, National Astronomical Observatories, Chinese Academy of Sciences, Beijing 100101, China\\
        \and
          School of Astronomy and Space Sciences, University of Chinese Academy of Sciences, Beijing 100049, China\\
        \and
          CSIRO Astronomy and Space Science, ATNF, Box 76 Epping, NSW 1710,
Australia\\
        \and Kavli Institute for Astronomy and Astrophysics, Peking University, Beijing 100871,
China
   }

\date{Received~~2019 xx xx; accepted~~20xx xx xx}

\abstract{The fast radio burst, FRB~171019, was relatively bright when discovered first by ASKAP, but was identified as a repeater with three faint bursts detected later by GBT and CHIME. These observations lead to the discussion of whether the first bright burst shares the same mechanism with the following repeating bursts. A model of binary neutron star merger is proposed for FRB 171019, in which the first bright burst occurred during the merger event, while the subsequent repeating bursts are starquake-induced, and generally fainter, as the energy release rate for the starquakes can hardly exceed that of the catastrophic merger event. This scenario is consistent with the observation that no burst detected is as bright as the first one.
\keywords{pulsars: general -- stars: neutron -- dense matter --
gravitational waves}
  }
  \maketitle

\section{Introduction}

Fast radio bursts (FRBs), are millisecond extragalactic radio flashes, still have mysteries on their cosmological origins \citep{Lorimer07,Keane12,Thornton13,Kulkarni14,Petroff15,Petroff16,Chatterjee17}.
Up until now, dozens of FRBs have been identified as repeaters \citep{Spitler16,CHIME19a,CHIME19b}.

A very interesting open question is whether all FRBs repeat.
There are a lot of efforts having been made to study the relationship between FRB repeaters and non-repeating FRBs.
The first repeater, FRB 121102, was localized in a low-metallicity dwarf ($\sim10^8\,M_{\odot}$) starforming galaxy at a redshift of $z=0.193$ \citep{Chatterjee17,Marcote17,Ten17} with an extremely magneto ionic environment \citep{Michilli18}, which directly confirmed the cosmological origin.
However, recently, a single FRB which has not been observed repetition, was localized in a more massive spiral galaxy \citep{Bannister19}, in contrast to the host galaxy of FRB 121102.
The differences of their host galaxy lead to the discussion of multi-origins between repeaters and non-repeating FRBs.

Repeaters also exhibit some different properties with non-repeating FRBs.
For instance, the time--frequency downward drifting pattern appears in at least some of repeaters' sub-pulses \citep{Hessels19,CHIME19a,CHIME19b}, while non-repeating FRBs lack such structures, suggesting that these are most likely to be a common feature for repeaters \citep{Wang19}.
Additionally, repeating sources tend to show less luminous bursts than most non-repeating FRBs \citep{Luo18,Luo19}.
Both above may hint that, possibly, repeating FRBs share different radiative and energy-providing mechanism with non-repeating FRBs.

\begin{figure}
\begin{center}
\includegraphics[width=0.8\textwidth]{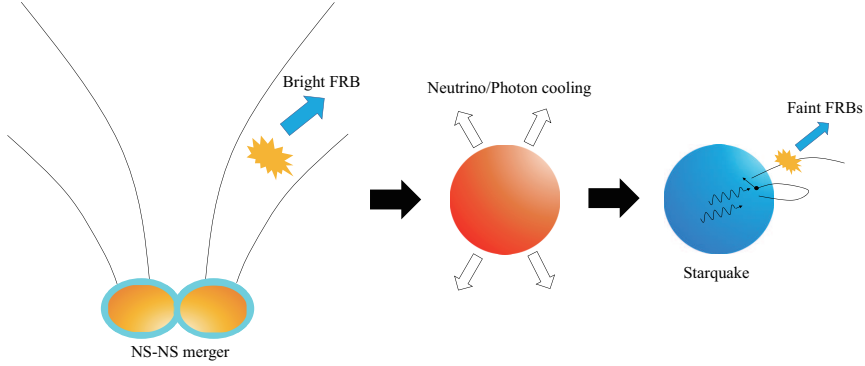}
\caption{\small{%
Schematic diagram of the model. The first discovered bright FRB is supported by fast orbital energy release during the two NSs merger. After that, a post-merger of long-lived neutron star forms and cools rapidly because of the neutrino/photon cooling. Starquakes would occur after the stellar solidification. The magnetic line footpoint oscillation may re-activate magnetospheric $e^\pm$-pair production or even trigger magnetic reconnection, which accelerates charged bunches emitting the following faint FRBs.
}} \label{f1}%
\end{center}
\end{figure}

Motivated by the recent observation of FRB 171019, which was
reported to exhibit repeating bursts $\sim590$ times fainter than
its discovery burst \citep{Kumar19}, we propose a FRB engine of merging normal neutron star (NS) or strangeon star (SS, see \citealt{Xu18} for a brief introduction) to power both non-repeating catastrophic and repeating bursts (see Fig.~\ref{f1}).
In fact, several catastrophic events, which can generate new-born pulsar-like
compact stars, have already been involved to interpret non-repeating bright FRBs
(e.g., \citealt{Jin18,Kashiyama13,Totani13,Wang16,Liu17}). 
\cite{Yamasaki18} speculated that the remnant pulsar after the merger event could reproduce periodically repeating FRBs.
The following repeating FRBs could be caused by the stellar rigidity-induced starquakes
\citep{Wang18,Suvorov19} on the remnant compact star (either normal neutron
star or stangeon star).
The model is introduced in Section \ref{sec:model}, and its applications to FRB~171019 are discussed in Section \ref{sec:frb}. Summary and discussion are in Section \ref{sec:summary}.

\section{The Model}\label{sec:model}
\subsection{Non-repeating FRB generated by NS-NS or SS-SS merger}

From the observations of many non-repeating FRBs, one can estimate the luminosity
\begin{equation}
L=4\pi D^2\frac{\delta \Omega}{4\pi}S_{\nu}\Delta \nu\simeq10^{43}D_{\rm Gpc}^2\left(\frac{\delta \Omega}{4\pi}\right)\left(\frac{S_{\nu}}{10\,\rm Jy}\right)\Delta\nu_{\rm GHz}~~\mathrm{erg\,s^{-1}},
\label{eq1}
\end{equation}
where $D_{\rm Gpc}$ is the source distance in Gpc, $\delta \Omega$ is the beam angle, $S_\nu$ is the flux density and $\Delta \nu_{\rm GHz}$ is the frequency bandwidth in GHz.
Such bright emission is generally considered to be coherent radiation.

\cite{Totani13}, \cite{Wang16} and \cite{Murase17} concluded that FRB emission can occur during the precursor phase of compact star mergers. In this model, we propose that the one-off FRBs are generated by NS-NS or SS-SS merger. Our model is based on the unipolar inductor model, which was first proposed for the Jupiter-Io system~\citep{GL}. The event rate of the merger bursts should be equivalent to the NS-NS merger rate, viz. $\sim 10^{-1}-10^3\mathrm{Gpc^{-3}\,yr^{-1}}$ (\citealt{Chruslinska18,Paul18,PML19}).

The orbital angular velocity can be calculated as $\Omega=(GM(1+q)/a^3)^{0.5}\sim3.7\times10^3\,\rm rad\,s^{-1}$, where $q=1$ is adopted as the mass ratio of the two NSs, $a=30\,\rm km$ is adopted as the distance between two NSs and $M=1.4\mathrm{M_{\sun}}$ adopted as the mass of NS. In the late inspiral of binary coalescence, the dipolar configuration of the magnetic field between two NSs should be severely distorted due to magnetic interaction, while the far field should remain dipolar with a total magnetic moment ${\vec \mu}_\mathrm{tot}={\vec\mu}_1+{\vec\mu}_2$. Inside the region between the two NSs, charged particles are able to move across the orbital plane in the distorted magnetic field, therefore the magnetosphere around the orbital plane can be treated as a conducting plate. Under the assumption of magnetic moment conservation and a parallel configuration, the surface magnetic field can be estimated as
\be
B\sim\frac{\mu_{\mathrm{tot}}}{\pi a^{2}R_{*}}\sim \frac{8}{3}\left(\frac{R_*}{a}\right)^2B_*=3.0\times 10^{11}B_{*, 12}R_{*,6}^2a_{6.5}^{-2}\mathrm{G},
\ee
where $a\sim 30\mathrm{km}$, $R_*\sim 10\mathrm{km}$ is the typical pulsar radius, $B_*\sim 10^{12}\mathrm G$ is the typical surface magnetic field on pulsars, and the convention $Q _n= Q/10^n$ in cgs units is adopted.

The
orbital evolution of the binary is dominated by gravitational-wave
radiation. Therefore, one can estimate the distance evolution
as~\citep{Peters64}
\be
a=a_0\left[1-\frac{256G^3M^3q(1+q)}{5a_0^4c^5}t\right]^{\frac{1}{4}}=20(1-1695t)^{\frac{1}{4}}~{\rm
km}.
\ee %
We set $a_0=20$ km for the case when the surfaces of the two NSs
touch with each other. One can then estimate the timescale of $\sim
2$ ms for the process (from $a=30$ km to $a_0=20$ km for $q=1$),
which is consistent with the FRB duration.

 On the conducting plate, the open field line region can be estimated by the proportion of its magnetic flux against the total magnetic flux,
\be
r_{cap}=(a+R_*)\sqrt{\frac{\Phi_\mathrm{open}}{\Phi_\mathrm{tot}}}=(a+R_*)\left(\frac{\int_{R_\mathrm{lc}}^{+\infty}\frac{r}{r^3}\mathrm dr}{\int_{a+R_*}^{+\infty}\frac{r}{r^3}\mathrm dr}\right)^{\frac{1}{2}}=(a+R_*)^{\frac{3}{2}}R_\mathrm{lc}^{-\frac{1}{2}},
\ee
where $R_\mathrm{lc}=c/\Omega$ is the radius of light cylinder.

A potential drop produced in the unipolar model is
\be
U\simeq\frac{B\Omega r_{\rm cap}^2}{2c}
\ee
This potential can trigger pair-production avalanches that creates charged bunches.
The bunches stream outward along open magnetic field lines, generating coherent radio emissions.
The  charge density in the magnetosphere of a pulsar is
given by \citep{GJ}
\be
\rho_e\simeq\rho_{\rm GJ}\simeq\frac{\Omega B}{2\pi c}
=5.8\times 10^3M_{33.44}^{\frac{1}{2}}R_{*,6}^2B_{*,12}a_{6.5}^{-\frac{7}{2}}\left(\frac{1+q}{2}\right)^{\frac{1}{2}}\mathrm{esu},
\ee
where  $\rho_\mathrm{GJ}$ is the Goldreich--Julian density, $B$ is
the magnetic field. The energy releasing rate of the magnetosphere during merger is given by
\be
\begin{aligned}
\dot E&\simeq U\pi r_\mathrm{cap}^2\rho_e c\simeq\frac{16}{9c^3}B_*^2G^2M^2(1+q)^2\left(\frac{R_*}{a}\right)^4\left(1+\frac{R_*}{a}\right)^6\\
&=6.3\times 10^{44}B_{*,12}^2M_{33.44}^2\left(\frac{1+q}{2}\right)^2\left(\frac{R_{*,6}}{a_{6.5}}\right)^4\left(1+\frac{R_{*,6}}{a_{6.5}}\right)^6\mathrm{erg\,s^{-1}}
\label{eq:L_oneoff}
\end{aligned}
\ee
If radio efficiency is $\sim 10^{-3}$,the beaming factor is $10^{2}$, the  isotropic luminosity matches the typical luminosity of non-repeating FRBs,  $L_\mathrm{iso}\sim 10^{-3}\times 10^{2}\times\dot E\sim 10^{43}\mathrm{erg/s}$.

The above calculation assumed the magnetic moments of the two merger stars are both parallel to the normal direction of the orbital plane (usually denoted as the ``U/U case"), which should result in the largest $\dot E$. The above result is consistent with the $\dot E$ of the U/U case in \citealt{Palenzuela13, Ponce14, WangJS18}.  For oblique configurations, the total magnetic moment $|\vec\mu_\mathrm{tot}|=|\vec\mu_1+\vec\mu_2|<|\vec\mu_1|+|\vec\mu_2|$. And the antiparallel configuration results in the smallest  $|\vec\mu_\mathrm{tot}|=|\vec\mu_1-\vec\mu_2|$, which can be smaller than $\mu_1$ and $\mu_2$ by several orders of magnitudes.

The compact star mergers may also produce other observable effects. The detection of GRB 170817A corresponding to GW 170817 \citep{Abbott17}  indicates that short $\gamma$-ray bursts can be generated by compact star mergers. Therefore, our model infers that a short GRB may be generated as the counterpart of a one-off FRB, which is consistent with the $\gamma$-ray counterpart to FRB 131104 \citep{DeLaunay16}.

\subsection{Subsequent repeating FRBs generated by starquakes}

We predict a new-born pulsar-like compact star left after a fraction of NS-NS or SS-SS merger. We suggest that the remnant object should live for at least several years to generate the subsequent repeating FRBs. NS-NS or SS-SS mergers do not always deliver long-lived compact stars. If the remnant mass is larger than the maximum mass for NS or SS, the merger product should be a blackhole.  With uniform rotation, the maximum mass of NS or SS can be larger than the Tolman-Oppenheimer-Volkoff (TOV) maximum mass, hence supermassive. And a supermassive remnant NS or SS, collapsing years after merger due to the braking of magnetic dipolar radiation or gravitational wave radiation, may also fit in our model. In addition, the geometric factor of the anisotropy further reduce the event rate of such sources. When taking the theoretical approach, the volumetric density repeating FRB sources is affected by the star formation history, the mass function of NS binaries, the merger dynamics and maximum mass of NS, some of which are still highly uncertain in theories. On the other hand, taking the observational approach, \cite{James19} limits the volumetric density of repeating FRBs to be $<70\mathrm{Gpc^{-3}}$. Adopting the burst rate of repeating FRB 121102, $5.7_{-2.0}^{+3.0}\,\mathrm{day^{-1}}$ \citep{Oppermann18}, as the burst rate for all repeating FRBs, the observed repeating bursts from such NS-NS remnants should be $\lesssim 10^5\,\mathrm{Gpc^{-3}yr^{-1}}$ for the more sensitive telescopes like GBT and CHIME. In this scenario, it’s rather difficult for ASKAP-like small dishes to detect repeating events, which is consistent with the prediction by \cite{Yamasaki18}.

When the initial proto-compact star is formed, gravitational energy
would be stored in the compact star as a form of initial thermal
energy. Consequently, the temperature of the inner core will be
$\sim (30-50)$ MeV. In the first following stage, neutrino emissions
make the star becoming cooling very rapidly. However, different
equation of states can lead to different cooling behavior.

On the one hand, in the regime of normal neutron star, for a
proto-neutron star, it shrinks into $\sim10$ km within several tens of
seconds because of the powerful neutrino-induced cooling down
\citep{Pons99}. The crust has a relatively lower neutrino emissivity
than the core. It makes the crust cools slower than the core and the
surface temperature decreases slowly during the first ten to hundred
years \citep{Chamel08}. After that, when the cooling wave from the
core reaches the surface, the surface temperature will drops
sharply.

On the other hand, the cooling process for a new-born strangeon star
(SS) consists of three stages \citep{Dai11}. The first stage is a
rapidly cooling process caused by the neutrino and photons emitting
at the very beginning. This process spends several tens seconds but
in principle faster than the first stage of an NS cooling
\citep{Yuan17}. The SS enters then the second stage, at which the
surface temperature remains constant roughly and the liquid SS
begins to be solidified, when the temperature drops to the melting
point temperature $T_p\sim0.1$\,MeV. The time scale of the liquid-solid
phase transition can be estimated as 
\be 
t_{\rm solid}=\frac{E_{\rm
in}}{4\pi R^2\sigma T_p^4}=7.8\times10^6E_{\rm in, 52}R_6^{-4} \,\rm s,
\label{eq2} 
\ee 
where $E_{\rm in}$ is energy release during the
phase transition, $\sigma$ is the Stefan-Boltzman constant and $R$
is the stellar radius. After the solidification, the newly-born SS
will rapidly release its residual inner energy because of its low
thermal capacity~\citep{YX}.

Basically, whether it is an NS or an SS, the magnetic field lines
are anchored to the stellar surface and their geometry is determined
by the motion of the footpoints. For a new-born compact star, there
are several kinds of instability which may be driven by
gravitational, magnetic or rotational energy. If the instability can
grow very fast in the stellar crust, making the pressure exceed a
threshold stress, crust quake would happen. Seismic waves, created
by the sudden release of energy, diffuse in the crust, which can help
some local instabilities growing. The characteristics of
self-organized criticality observed in earthquakes, are very
expected to be seen in some compact star activities (e.g.,
\citealt{Cheng96,Duncan98,Gog99}). The compact star, which is
suggested to be a dead pulsar (i.e., beyond the pulsar death line),
can be then excited due to the solid crust quake.
A similar but different story of strangeon
star was presented in \cite{Lin15}.

In the regime of normal NS, crust shear can trigger the footpoint
motion, therefore the magnetic curl or twist are ejected into the
magnetosphere from the crust in $\tau\simeq R/v_{\rm A}\sim1$\,ms
\citep{Thompson02}, where $v_{\rm A}$ is the Alv{\'e}n speed.
During this process, charged particles in the magnetosphere, are suddenly accelerated to be ultra-relativistic by the quake-induced magnetic reconnection, and form charged bunches.
In general, the cooling timescale for curvature radiation in the observer's rest frame is much smaller than FRB's typical duration.
Thus, it requires a strong electric field parallel $E_{\parallel}$ to the B-field that can accelerate electrons, supplying the kinetic energy to balance radiation loss, which is given by \citep{Kumar17}
\be
E_{\parallel}=\frac{\gamma_e mc}{et_{\rm cool}}=3\times10^8\nu_9N_{e,24}\gamma^{-2}_2\,\rm esu,
\ee
where $\gamma_e$ is the Lorentz factor of electrons, $\nu$ is the emission frequency of curvature radiation, and $N_e$ is the electron number.
The $E_{\parallel}$ to the B-field formed by the magnetic reconnection is given by
\be
E_{\parallel}\simeq\frac{2\pi \sigma_sv_{\rm A}B}{c}=2.1\times10^9\sigma_{s,-3}v_{\rm A, 8}B_{14}\,\rm esu,
\ee
where $\sigma_s=\xi/\lambda$ is the strain, in which $\xi$ is the amplitude of oscillations and $\lambda$ is the characteristic wavelength of oscillations.

Let us consider the pair production forming charged bunches which emit coherent curvature radiation.
Basically, several authors have discussed curvature radiation from charged bunches from pulsar magnetospheres to explain FRBs (e.g., \citealt{Katz14,Kumar17,Yang18}).
If the charge density prompted by the stellar rotation $\rho_{\rm GJ}$ and local twist is insufficient to screen $E_{\parallel}$, sparks would occurs at the polar gap regions which creates emitting charged bunches \citep{Wad19}.
The number density due to the footpoint motion can be estimated as
\be
n_e\simeq\frac{E_{\parallel}}{4\pi e\lambda}=3.5\times10^{12}\sigma_{s,-3}\Omega_{\rm osc,3}B_{14}\,\rm cm^{-3},
\ee
where $\Omega_{\rm osc}$ is the oscillation frequency.
Here we assume that the seismic wave is dominated by the base mode which $\Omega_{\rm osc}\sim(\pi/R)(\mu_s/\rho_n)^{0.5}\sim10^3\,\rm Hz$, where $\mu_s$ is the shear modulus and $\rho_n$ the neutron drip density.
To generate coherent emissions, charged particles in the bunch would emit with approximately the same phase.
Therefore, a comoving volume of the bunch can be estimated as $\eta(\gamma_e c/\nu)^3$.
Only fluctuating electron can contribute to the coherent radiation, the number of which is given by
\be
N_e=\mu n_e(\frac{\gamma_e c}{\nu})^3=9.4\times10^{23}\mu\eta_1\sigma_{s,-3}\Omega_{\rm osc,3}B_{14}\gamma_{e,2}^3\nu_9^{-3},
\ee
where $\mu$ is the fraction of electrons fluctuation and $\eta$ is the multiplication factors due to the frame transform \citep{Kumar17}.
The total luminosity of the curvature radiation from charged bunches can be described as $L_{\rm iso}=N_{\rm pat}(N_e^2\delta L_{\rm iso})$, where $\delta L_{\rm iso}=2e^2\gamma_e^8c/(3\rho^2)$ is the isotropic luminosity in the observer frame, and $N_{\rm pat}$ is the patch number.
Thus, we have
\be
L_{\rm iso}=8.3\times10^{39}N_{\rm pat}\mu_{-1}^2\eta_1^2\sigma_{s,-3}^2\Omega_{\rm osc,3}^2B_{14}^2\gamma_{e,2}^8\nu_9^{-4}\,\rm erg\,s^{-1}.
\label{eq:L_starquake}
\ee
The calculated luminosity is consistent with observations of most repeaters.

The energy release rate during the magnetic reconnection and
oscillation-driven activity can both be estimated as \be {\dot
E}_{\rm R}\simeq4\pi R_{p}^2\delta R\frac{\sigma_sB^2}{8\pi
\tau}=3.3\times10^{42}P_{-3}^{-1}\delta
R_4\sigma_{s,-3}B_{14}^2\tau^{-1}_{-3}\,\rm erg\,s^{-1}, \ee where
$R_p$ is the polar cap radius of the remnant star and $\delta R$ is the height of the
crust. The energy release rate for the following starquakes is much
smaller than that of the merger stage. Therefore, one expects that
there is no any burst detected in the follow-up observation to be as bright as the one triggered by NS-NS
merger.

\section{Properties of FRB 171019}\label{sec:frb}
FRB 171019 was originally detected in a wide-field survey of ASKAP \citep{Shannon18}, hereafter referred to as the "ASKAP burst". Two weaker repetitions of FRB 171019 were detected in GBT searches on 2018/07/20 and 2019/06/09 \citep{Kumar19}, hereafter referred to as "GBT bursts". Another repetition was detected by CHIME on 2019/08/05 \citep{Patek19}, hereafter referred to as the "CHIME burst". Some properties of the bursts are shown in Table \ref{tab:properties}.

\begin{table}[!hbt]
\centering
\begin{threeparttable}
\caption{Properties of FRB 171019 and its repetitions.}
\label{tab:properties}
\begin{tabular}{c c c c c c c c }
\toprule
\multirow{2}{*}{No}&\multirow{2}{*}{Telescope}&Obs Freq&Obs Time&TOA\tnote{d}&DM &Fluence&Burst Width\\
&&($\mathrm{MHz}$)&($\mathrm h$)&(MJD)&($\mathrm{pc/cm^3}$)&($\mathrm{Jy\cdot ms}$)&($\mathrm{ms}$)\\
\midrule
1\tnote{ab}&ASKAP&1129.5-1465.5&$986.6$&$58045.56061371$&$461\pm1$&$219\pm5$&$5.4\pm0.3$\\
\midrule
2\tnote{b}&\multirow{2}{*}{GBT}&\multirow{2}{*}{720-920}&\multirow{2}{*}{$10.6$}&$58319.356770492$&$456.1\pm0.4$&$0.60\pm0.04$&$4.0\pm0.3$\\
3\tnote{b}&&&&$58643.321088777$&$457\pm1$&$0.37\pm0.05$&$5.2\pm0.8$\\
\midrule
4\tnote{c}&CHIME&400-800&$17\pm 3$&$58700.38968$&$460.4\pm0.2$&$\gtrsim 7$&$6\pm2$\\
\bottomrule
\end{tabular}
\begin{tablenotes}
\item[a] \cite{Shannon18}
\item[b] \cite{Kumar19}
\item[c] \cite{Patek19}
\item[d] Burst time of arrivals are referenced at different frequencies: 1464 MHz for ASKAP, 920 MHz for GBT, and 400 MHz for CHIME.
\end{tablenotes}
\end{threeparttable}
\end{table}

The ASKAP burst is $\sim 600$ times brighter than the GBT bursts. It is highly possible that the ASKAP burst and the GBT bursts belong to two separate classes of FRBs. In the detailed model calculated in Section \ref{sec:model}, we propose that the initial bright burst is generated in a catastrophic event, viz. merger of compact star binary, whereas the two much fainter repetitions are caused by repeating mechanism, viz. starquakes on the merger remnant star.

\begin{figure}
\begin{center}
\includegraphics[width=0.6\textwidth]{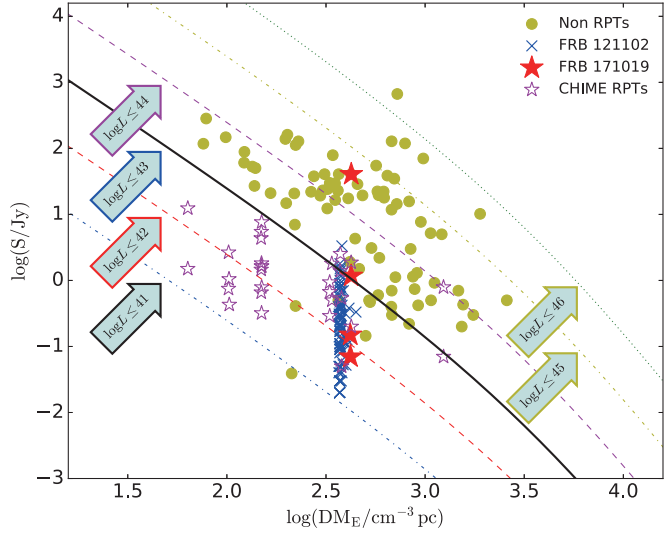}
\caption{\small{%
A radio flux-dispersion measure ($S-{\rm DM_E}$) diagram for FRBs,
with ${\rm DM_E}$ the dispersion measure deducting the contribution
by the Milk Way at the direction. Dots are for non-repeating FRBs,
filling pentagrams for FRB~171019, empty pentagrams for CHINE
repeaters, and cross marks for FRB~121102.
The curves represent approximately constant radio luminosity.
}} \label{f2}
\end{center}
\end{figure}

The the luminosity of a merger-induced burst given by Eq\eqref{eq:L_oneoff} is $\sim 10^3$ times larger than the luminosity of starquake-induced bursts derived in Eq\eqref{eq:L_starquake}. It is noteworthy that this ratio is consistent with the brightness difference between the ASKAP burst and the GBT bursts. In addition, the luminosity of non-repeating bursts given by Eq\eqref{eq:L_oneoff} lies in the more luminous region in Fig.\ref{f2}, while the luminosity of starquake bursts in Eq\eqref{eq:L_starquake}  lies in the less luminous region. The CHIME burst is $\sim 10$ times brighter than the GBT busts, which may be explained by an increase of the patch number $N_{\mathrm{pat}}$ in our model.

Another noteworthy fact is the non-detection of Parkes follow-up observation. According to \cite{Kumar19}, the Parkes 64-m telescope observed the source for 12.4 hours in total during the 8 month succeeding the ASKAP detection, however, no repeating burst was detected even though the Parkes telescope is much more sensitive than ASKAP. This time length is longer than the timescale given by Eq\eqref{eq2}, therefore it is consistent with the timescale of solidification and stress accumulation on the new-born compact star.

However, it is still possible that this luminosity difference derived from observation is caused by the selection effect of observations. Limited by the relatively lower sensitivity, a burst as bright as the GBT bursts or the CHIME burst should be undetectable at ASKAP in fly's eye mode. Though the GBT and the CHIME observations did not detect bursts as bright as the ASKAP one, it must be taken into consideration that the total observation time at GBT is only 10.6 hours, and only $17\pm 3$ hours at CHIME, which are much shorter than the 986.6 hour observation using ASKAP. If so, there could be alternative explanations, e.g. the bursts are generated by a unified mechanism with a intensity distribution, and the follow-up observations are too short to detect bright ones. In the future, should a repetition burst as bright as FRB 171019 occur, our ``merger+starquake" model could be ruled out.

For the binary merger involved in our explanation, it should be an NS-NS or SS-SS merger resulting in a long-lived NS/SS afterwards, rather than a WD-WD or blackhole-involved merger. A WD-WD merger would generate a type Ia supernova, however, no optical counterpart of FRBs has been detected. In addition, the supernova remnant would be optically thick for radio emission in a long time after the merger, which is in contradiction with the FRBs detected. As for blackhole-involved models, the product would be a blackhole, which cannot explain the repeating bursts.

After the NS-NS merger, the mass ejection will shield the following radio emission of the nascent pulsar.
Since the optical depth of the relativistic jet launched after the merger decays faster than the slower ejection, and the kilonova ejecta is optically thick in L-band within several decades succeeding the merger \citep{Margalit18}, the repeating FRBs should be along the relativistic jet,
such that one can detected these repeating FRBs under our model.
But electrons in the relativistic jet may result in the difference of $\rm DM$ between the one-off FRB and repeating FRBs.
The total number of electrons in the relativistic jet can be estimated as
\begin{eqnarray}
N_{\rm e}\sim \frac{E_{\rm jet}}{\Gamma m_{\rm p}c^{2}}\sim \pi(r\theta )^{2}l n_{\rm e},
\end{eqnarray}
where
\begin{eqnarray}
r\sim c\Delta t
\end{eqnarray}
is the distance from the relativistic jet to source, $\Delta t$ is time interval between the one-off FRB and repeating FRBs,
$E_{\rm jet}$ is total energy of the relativistic jet, $n_{\rm e}$ is number density of electrons, $\Gamma$ is saturated bulk Lorentz factor of the relativistic jet,
$m_{\rm p}$ is proton mass, $\theta$ is jet opening angle, $l$ is thickness of the relativistic jet.
Therefore, the contribution of relativistic jet to $\rm DM$ is
\begin{eqnarray}
\mathrm{DM}&=&\int _{r}^{r+l}n_{e}(r')dr'\sim\frac{E_{\rm jet}}{\pi \Gamma m_{\rm p}c^{4}\Delta t^{2}\theta^{2} }\nonumber\\
&=&7.8\times 10^{-3}E_{\rm jet,50}\Gamma_2^{-1}\theta_{-1}^{-2}\Delta t_{7}^{-2}\;\rm pc\cdot cm^{-3}.
\end{eqnarray}
It is clear that the effect of relativistic jet on $\rm DM$ is negligible. However, several decades after the merger, the kilonova ejecta would become transparent and contribute $\sim 10^2\,\mathrm{cm^{-3}pc}$ to the total DM \citep{Margalit18}, which may be observable for some sources but not in the case of FRB 171019. It is worth noting that the pulsar of post-merger could be spin down so significantly that it's radio luminosity can hardly be detectable on the Earth unless enhanced by starquakes.

\section{Summary and Discussion}\label{sec:summary}
Observations show the first brighter burst of FRB 171019 followed by three weaker repeaters about one year later.
In this paper, we propose a unified frame to explain this feature.
(i) The first one-off FRB is generated at the moment before NS-NS or SS-SS merger through, e.g., unipolar inductor mechanism.
(ii)The nascent remnant SS takes $\sim 100\;\rm d$ to be solidified (see equation \ref{eq2}) which accounts for the halcyon period between the one-off burst and the followed repeaters.
(iii) After the solidification, starquakes induced by the spin-down of the SS generate the subsequent three weaker repeating FRBs.

Although, the event rate of NS-NS/SS-SS merger seems much smaller than that of FRBs, the jump feature on the luminosities of the
first brighter burst of FRB 171019 and the followed repetitions indicates this repeating FRB may belong to a special subclass.
Especially, there is probability to directly test our model since the intense one-off burst has a different mechanism from the weak repetitions.
For example, one can keep monitoring this source, if another bright burst just like the first one of FRB 171019 were to be detected, it would mean that our model should be ruled out.

Can a massive NS survive the merger event of binary NSs? This is really a problem essentially related to the fundamental strong
interaction at low-energy scale and hence to the non-perturbative quantum
chromo-dynamics, the equation of sate of cold super-dense matter, which still
remains a challenge. Nevertheless, it is generally thought that strangeness would
play an important role in understanding the state of bulk strong matter \citep{Xu18}, and pulsars should not be conventional NSs but strangeon stars (SSs),
formerly called quark cluster star \citep{Xu03}. The equation of state of
strangeon matter would be so stiff
that its maximum mass could be as high as $\sim 3M_\odot$ \citep{Lai09}, and the
later discoveries of $2M_\odot$-pulsars fit the expectation. Furthermore, the
merger event of GW~170817 can also be well understood in the SS model \citep{Lai19,Baiotti19}. Therefore, we anticipate that the
unknown equation of state could be the first big problem to be solved in the era of gravitational-wave
astronomy.

\section*{Acknowledgements}

The authors thank Dr. Bing Zhang for his valuable discussions. This
work is support by MoST Grant (2016YFE0100300), National Key R\&D
Program of China (2017YFA0402602), NSFC (11633004, 11473044,
11653003, 11673002 and U1531243), the Strategic Priority Research
Program of CAS (XDB23010200), and the CAS grants (QYZDJ-SSW-SLH017
and CAS XDB 23040100).

\bibliographystyle{mnras}
\bibliography{reference}

\end{document}